\documentclass[conference]{IEEEtran}
\IEEEoverridecommandlockouts

\usepackage{cite}
\usepackage{amsmath,amssymb,amsfonts}
\usepackage{algorithmic}
\usepackage{hyperref}
\usepackage{graphicx}
\usepackage{multirow}
\usepackage{bbding}
\usepackage{textcomp}
\usepackage{makecell}
\usepackage{array}
\usepackage{booktabs}
\usepackage{pgfmath}
\usepackage{utfsym}
\usepackage{fontawesome}
\usepackage{float}
\usepackage{url}
\usepackage[numbers,sort&compress]{natbib}

\usepackage{colortbl}
\definecolor{modelblue}{RGB}{220, 230, 255}      % GPT, GLM, Moshi, Minmo
\definecolor{modelgreen}{RGB}{220, 255, 220}     % LLaMA-Omni2 series
\definecolor{modelorange}{RGB}{255, 240, 220}    % Phi-MM series

\usepackage{tabularx}
\usepackage{pifont}
\usepackage{subcaption}
\usepackage{comment}

\usepackage{pifont} % include some cool symbols
\definecolor{bsRed}{rgb}{0.95, 0.0, 0.0}

\def\BibTeX{{\rm B\kern-.05em{\sc i\kern-.025em b}\kern-.08em
    T\kern-.1667em\lower.7ex\hbox{E}\kern-.125emX}}
\begin{document}

\title{Towards Efficient Speech-Text Jointly Decoding within One Speech Language Model
% ESI-Omni: Early-Stop Interleaving for Efficient Speech Text Jointly Decoding in Speech Language Models
% Towards Efficient Alignment of Output Speech and Text in Speech Language Models
\thanks{* Equal contribution.}
}

\author{
Haibin Wu*, Yuxuan Hu*, Ruchao Fan, Xiaofei Wang, Kenichi Kumatani,
Bo Ren, Jianwei Yu, Heng Lu, \\ Lijuan Wang, Yao Qian, Jinyu Li
\\
\textit{Microsoft, USA}
}
% \author{Anonymous Author}

\maketitle

\begin{abstract}
Speech language models (Speech LMs) enable end-to-end speech-text modeling within a single model, offering a promising direction for spoken dialogue systems. 
The choice of speech-text jointly decoding paradigm plays a critical role in performance, efficiency, and alignment quality. 
In this work, we systematically compare representative joint speech-text decoding strategies—including the interleaved, and parallel generation paradigms—under a controlled experimental setup using the same base language model, speech tokenizer and training data. 
Our results show that the interleaved approach achieves the best alignment. However it suffers from slow inference due to long token sequence length. 
To address this, we propose a novel early-stop interleaved (ESI) pattern that not only significantly accelerates decoding but also yields slightly better performance. 
Additionally, we curate high-quality question answering (QA) datasets to further improve speech QA performance. 
% Our findings provide practical insights into decoding paradigm design for future Speech LMs.
\end{abstract}


\section{Introduction}
\label{sec:intro}

% Recent advances in large LMs (LLMs), e.g. ChatGPT~\cite{chatgpt}, have reshaped the way humans interact with machines, especially through text-based communication. However, real-world conversations are rarely limited to text alone—they also carry rich paralinguistic and acoustic details. 
% Voice-based interaction offers a more natural and intuitive interface, bringing with it elements like rhythm, emotion, and nuance that are difficult to capture through text. 
% One intuitive way for voice-based interaction is to rely on a pipeline~\cite{huang2024audiogpt} that cascades automatic speech recognition (ASR), text LM, and text-to-speech (TTS). Although this approach works in many cases, it often suffers from high latency, accumulated errors across components, and an inability to fully capture paralinguistic cues such as emotion and tone. 
% As a result, there is growing interest in speech LMs (Speech LMs)~\cite{arora2025landscape,defossez2024moshi,ji2024wavchat,wu2024towards,peng2024survey}, which aim to handle speech input and output in a fully end-to-end manner, enabling faster, more expressive, and more natural spoken interaction.

Speech language models (LMs)~\cite{arora2025landscape,defossez2024moshi,ji2024wavchat,wu2024towards,peng2024survey} are designed to generate both text and speech within a single model. Thus, the decoding paradigm—i.e., how the model represents and produces speech and text—is a critical design choice. This includes (1). how the model learns to jointly predict both modalities; and (1). how generated speech and text tokens are fed back into the model for next-token prediction.
Building effective speech LMs requires careful consideration of various factors, including pretraining design~\cite{arora2025landscape,defossez2024moshi,ji2024wavchat}, supervised fine-tuning (SFT) strategies~\cite{arora2025landscape,defossez2024moshi,ji2024wavchat}, speech tokenization~\cite{guo2025recent,shi2024espnet,wu2024codec,mousavi2024dasb}, data curation, and decoding paradigms~\cite{arora2025landscape,ji2024wavchat,defossez2024moshi}. 
In this paper, we focus mainly on the decoding paradigm and data curation for the SFT stage, as their success experience can be re-used and benefit other components of the system.

The prevailing decoding paradigms include:
(1). Interleaved decoding~\cite{nguyen2024spirit, zeng2024glm} alternates between generating text and speech tokens in a single sequence. At each step, the LM takes the previously generated token—either text or speech—as input for next-token prediction.
(2). Parallel decoding\cite{ding2025kimi, chen2024slam} predicts one text and multiple speech tokens in each forward pass. The averaged embedding of the generated text and speech tokens is fed back into the model for the next step. 
(3). Thinker-Talker~\cite{chen2025minmo, fang2025llama} separates text and speech generation: the Thinker (LLM) predicts text tokens, which are fed back into the model for next-token prediction, while the Talker generates speech tokens based on the Thinker’s hidden states and text outputs.
Previous studies often adopt different base LMs, speech tokens and datasets, making it difficult to conduct fair comparisons between various decoding paradigms.

\begin{figure*}[htp]
    \centering
    {\includegraphics[width=0.8\textwidth]{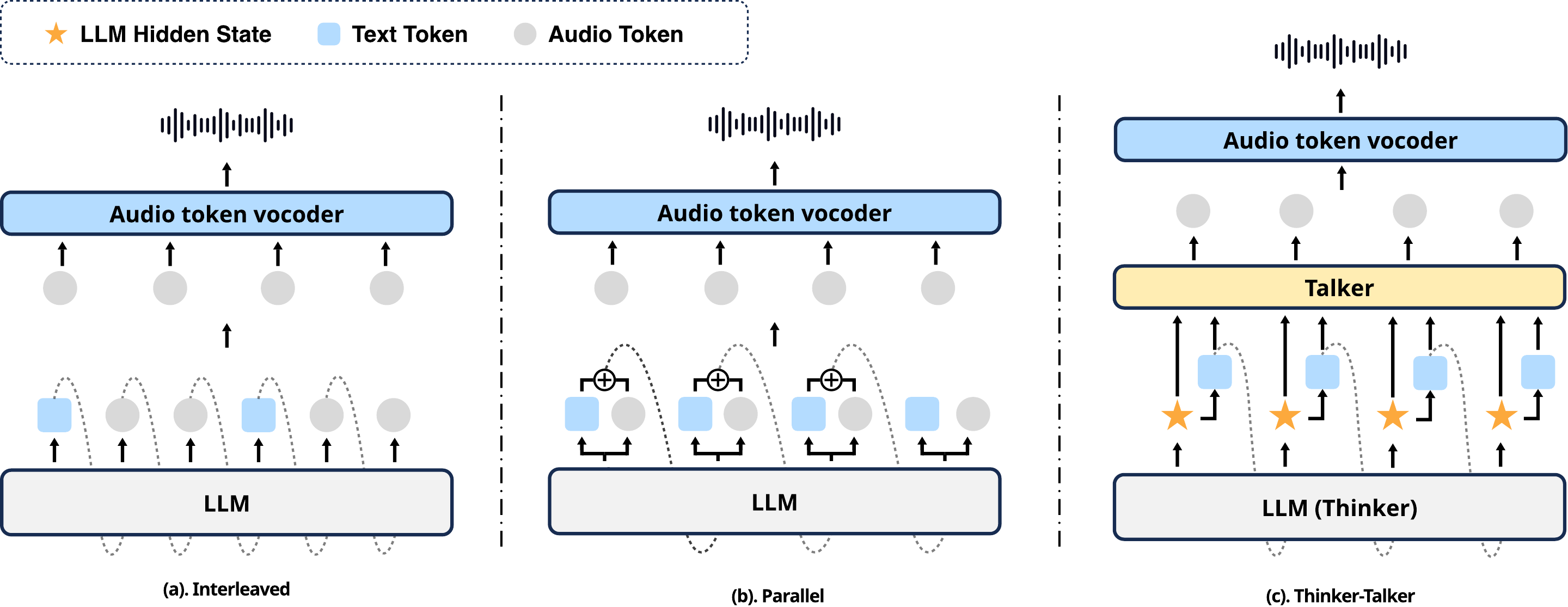}}
    \caption{Decoding patterns. (a). Interleaved pattern; (b). Parallel pattern; (c). Thinker-Talker pattern. (b) and (c) use LM to decode both text and speech tokens, while (c) uses the LM to decode only text tokens. We refer to \cite{chen2024slam} to draw the figure.}
    \label{figure:decoding_pattern}
    \vspace{-5pt}
\end{figure*}

The main three contributions of our work are the following:

\begin{itemize}
    \item \textbf{Fair comparison of decoding paradigms:} In this work, we systematically investigate and compare several representative decoding strategies for joint speech-text generation within one LM, including the interleaved and the parallel token generation approaches. To ensure a fair and controlled evaluation, we use the same base LM and training data across all settings.
    Our results show that the interleaved paradigm yields the most promising performance. Additionally, we observe that the Thinker-Talker approach requires careful tuning to achieve well-aligned speech and text outputs.
    \item \textbf{Accelerating interleaved decoding:} While the interleaved decoding paradigm achieves strong performance in joint speech-text modeling, its inference speed is often hindered by the long interleaved text-speech token sequences. To address this, we propose a novel early-stop interleaved (ESI) paradigm, where the model is trained to predict a special token following the end-of-sentence (EOS) text token. After this token is generated, the model produces only speech tokens, effectively skipping redundant text padding tokens and reducing the total sequence length to \textbf{75\%} of its original length. This approach significantly improves inference efficiency while achieving comparable or slightly better model’s intelligence and the alignment performance. % (The ratio of padding tokens to actual text tokens is around \textbf{2.95} to maintain a fixed interleaving ratio of 1 to 2). This approach significantly improves inference efficiency while achieving comparable or slightly better model’s intelligence and the alignment performance.
    \item \textbf{Data curation for speech QA:} We enhance the performance of speech-in, speech-out question answering by carefully curating and incorporating high-quality QA datasets. This curated data leads to a significant improvement in speech QA performance.
\end{itemize}
\section{Method}
\label{sec:method}

\subsection{Fair comparison of decoding paradigms}
\label{subsec:decoding_pattern_compare} 

We details three representative decoding paradigms, as illustrated in Figure~\ref{figure:decoding_pattern}:

\noindent
\textbf{Interleaved}~\cite{nguyen2024spirit,zeng2024glm} (Figure~\ref{figure:decoding_pattern}.(a)): In this paradigm, speech and text tokens are interleaved with a specific ratio (In Figure~\ref{figure:decoding_pattern}.(a), we use 1:2 as an example for illustration) within a single sequence. During generation, the language model alternates between producing text and audio tokens, and each generated token—whether speech or text—is fed back into the model for next-token prediction. 
This interleaved strategy has demonstrated strong performance for joint speech-text modeling, as shown in both prior work\cite{nguyen2024spirit,zeng2024glm,zeng2024scaling} and our experiments.
However, a major drawback of this approach is the excessive sequence length, especially when padding tokens are used to maintain a fixed text-to-speech interleaved ratio after text tokens are exhausted. This significantly slows down inference speed and reduces training efficiency.

\noindent
\textbf{Parallel}~\cite{defossez2024moshi,ding2025kimi,chen2024slam} (Figure~\ref{figure:decoding_pattern}.(b)): During each forward pass of the language model, one text token and multiple speech tokens (can be one or more speech tokens. For simplicity, we use one in Figure~\ref{figure:decoding_pattern}.(b) for illustration) are predicted. The resulting text and speech tokens are then averaged in the hidden space and sent back into the model for next-token prediction.
For example, SLAM-Omni~\cite{chen2024slam} uses the hidden states from the last transformer layer of the LM and applies a multilayer perceptron (MLP) to predict both text and audio tokens.

\noindent
\textbf{Thinker-Talker}~\cite{chen2025minmo, fang2025llama} (Figure~\ref{figure:decoding_pattern}.(c)): In this paradigm, the Thinker serves as the primary large language model responsible for predicting solely text tokens, while the Talker is an auto-regressive model that generates audio tokens based on the Thinker's hidden representations and predicted text tokens. Only the text tokens are fed back into the Thinker for next-token prediction.
Minmo~\cite{chen2025minmo} trains both Thinker and Talker using a mix of tasks, including ASR, answering speech questions, prediction of interruption, and others. In contrast, LLaMA-Omni 2~\cite{fang2025llama} is trained solely on speech question answering data, which helps isolate the impact of the decoding paradigm by reducing interference from other tasks when making comparisons.

\begin{table}[ht]
\centering
\caption{Language model input and output tokens of different decoding patterns for one forward pass.}
\label{tab:decoding_patterns}
\begin{tabular}{lcc}
\toprule
\textbf{Pattern} & \textbf{LM Input} & \textbf{LM Output} \\
\midrule
Interleaved      & Alternate speech / text & Alternate speech / text \\
Parallel         & Averaged speech+text  & Speech and text \\
Thinker-Talker   & Text            & Text  \\
\bottomrule
\end{tabular}
\end{table}

The comparison of different patterns are shown in Table~\ref{tab:decoding_patterns}.
Although decoding paradigms form the foundation of speech language models, comparing these strategies in a controlled setting remains challenging. Key difficulties include variations in training data, training recipes, base language model configurations, speech tokenization methods, and the number of trainable parameters.
To ensure a fair and meaningful comparison, we standardize most of critical components to compare the jointly speech-text modeling patterns within one language model, include the interleaved and the parallel patterns: using the same base language model, (Phi4- MM~\cite{phi4mini}), the same supervised fine-tuning (SFT) data, the same number of trainable parameters, and the same speech tokenization method (S3Tokenizer~\cite{du2024cosyvoice}).
In the Thinker-Talker decoding paradigm, the Talker is decoupled from the Thinker and functions similarly to a pseudo TTS module—taking text tokens and hidden states from the language model to generate speech tokens. This separation prevents speech token prediction from interfering with the Thinker’s behavior, thereby preserving the original language model’s text capabilities.
Since our primary focus is on jointly trained decoding paradigms within one LM, where the main language model both predicts and consumes text and speech tokens, we include the Thinker-Talker approach only as a reference for comparison~\cite{chen2025minmo, fang2025llama}.
We use the officially released checkpoints of LLaMA-Omni-2 for comparison, as its training setup closely aligns with ours—train on only a few thousand speech QA data, unlike other models trained on millions of diverse datasets. Additionally, LLaMA-Omni-2 was released recently, making it a timely and relevant baseline. 
We argue that it still serves as a reasonable baseline for comparing with other decoding patterns.

\begin{figure*}[htp]
    \centering
    {\includegraphics[width=0.8\textwidth]{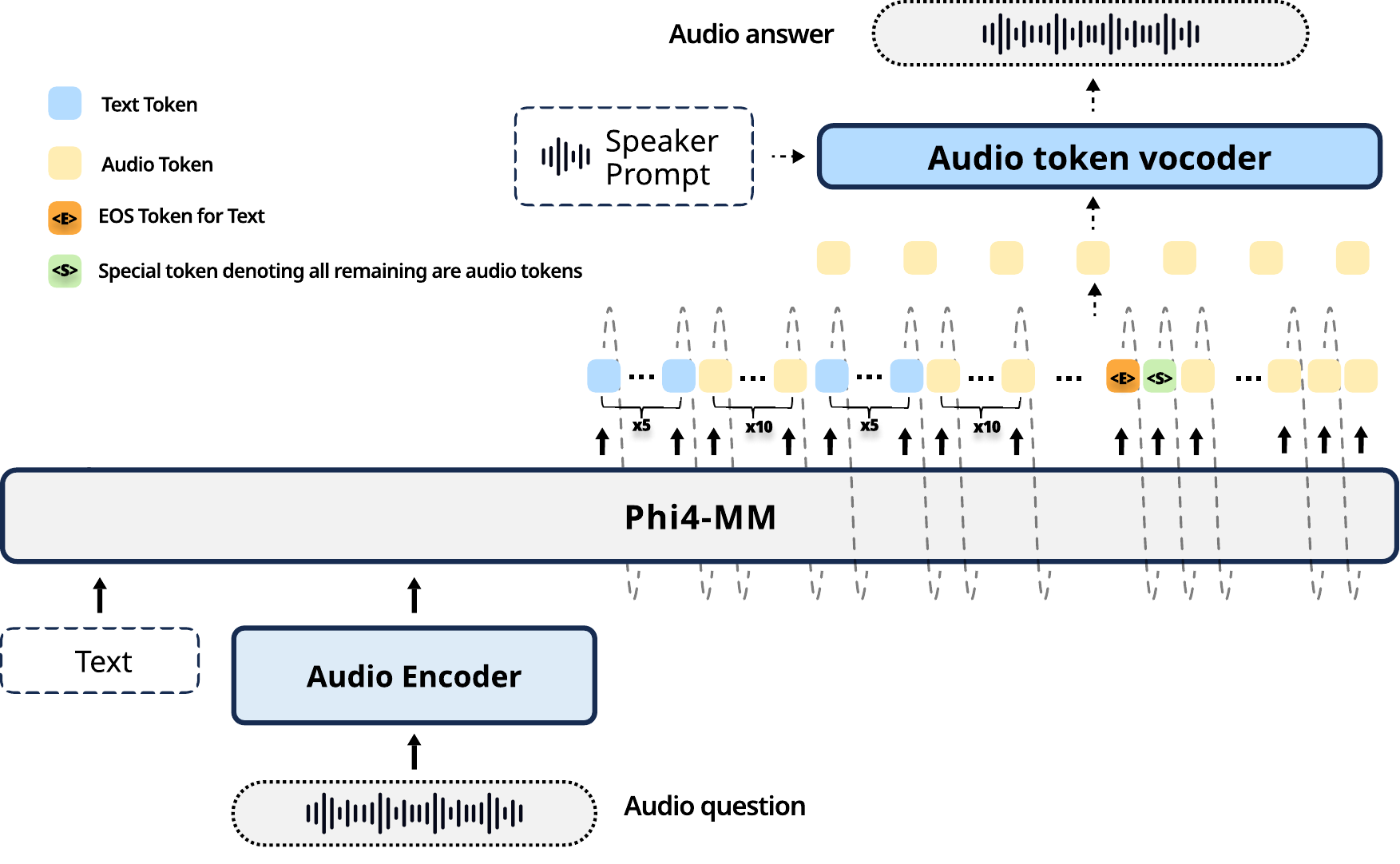}}
    \vspace{-10pt}
    \caption{The proposed early stop decoding paradigm to accelerate the interleaved pattern.}
    \vspace{-10pt}
    \label{figure:early_stop_pattern}
\end{figure*}

\subsection{Phi4-MM with ESI pattern}
\label{subsec:early_stop}

\subsubsection{What can be optimized to improve the efficiency of the interleaved pattern}
The interleaved decoding strategy offers effective joint modeling of speech and text, shown in previous papers and our experiments, but its practical deployment is often limited by slow inference due to lengthy interleaved text-speech token sequences.
For reference, a human typically speaks about 2–3 words per second, while the adopted speech tokenizer converts one second of audio into approximately 25 tokens. 
In a typical and high-performance interleaved setup with a text-to-speech token ratio of 1:2~\cite{zeng2024glm,chen2024slam} (as shown in Figure~\ref{figure:decoding_pattern}.(a)), text padding tokens are needed once the text tokens are exhausted, in order to maintain a consistent interleaving ratio throughout the sequence.
As a result, the language model ends up predicting a large number of text padding tokens\footnote{We randomly sample and then analyze 3,000 samples, and find that the ratio of padding tokens to actual text tokens is \textbf{2.95}.}, which consumes significant computation and increases inference latency.

\subsubsection{Proposed early-stop interleaved decoding paradigm}
To address the inefficiency caused by long interleaved sequences, as shown in Figure~\ref{figure:early_stop_pattern}, we introduce an early-stop interleaved decoding mechanism that enables the model to skip unnecessary text padding tokens. 
We construct interleaved sequences using a fixed text-to-speech token ratio of 5:10 before the text tokens are exhausted. 
Once the model finishes generating all meaningful text tokens and predicts the end-of-sentence (EOS) token for text, then a special token ($<\text{S}>$) is inserted to indicate that all remaining tokens in the sequence will be speech tokens\footnote{In our preliminary experiments, we observed that without the special token $<\text{S}>$, the language model struggled to initiate audio token generation. This token acts as a marker to signal the start of the remaining audio sequence, prompting the model to switch into a mode where all subsequent tokens are treated as audio tokens.}.
This design enables the removal of redundant text padding tokens that would otherwise be inserted to maintain a fixed interleaving ratio. For example, with a text-to-speech token ratio of 5:10, eliminating padding reduces the total sequence length to approximately 75\% of its original length, resulting in lower computational overhead and enhancing both inference speed and training efficiency.

Notably, removing text padding alters the original interleaving structure, resulting in a different decoding pattern for the latter part of the sequence. While this change might be expected to degrade performance, our experiments show a slight improvement instead. 
We hypothesize that adding too many padding tokens makes the token sequence unnecessarily long. During decoding, each token attends to all previous tokens, and as the sequence becomes longer, the model tends to focus less on the earlier, important tokens. Padding tokens make this problem worse, making it harder for the model to maintain attention on meaningful content at the beginning, which may lead to reduced performance.
Another possibility is that padding tokens, being less semantically meaningful compared with text tokens, and adding them may introduce noise and distract the model during decoding. 
Removing these padding tokens allows for shorter sequence length, cleaner and more focused generation.

\subsubsection{System overview of Phi-4MM with ESI}
Figure~\ref{figure:early_stop_pattern} presents an overview of our Phi4-MM with ESI, a speech-in, speech-out question answering system, built based on a multimodal open-source language model (Phi4-MM~\cite{phi4mini}).
The input text is consist of system prompts or user instructions. The audio question is first processed by an audio encoder—initialized from a pretrained ASR encoder and further fine-tuned on ASR data—to produce speech embeddings. These embeddings, together with the text input, are fed into the main speech language model.
We use the S3Tokenizer~\cite{du2024cosyvoice} to extract the speech tokens.
The model performs decoding using our proposed early-stop decoding pattern, generating both text and speech tokens. The generated text can be directly presented to the user as the textual answer, while the speech tokens are passed to a streaming audio token vocoder~\cite{du2024cosyvoice}, to synthesize the speech response.

\subsection{QA data curation}
\label{subsec:data_curation}

To enhance the language model’s ability in speech-to-speech question answering, we curate a speech question answering dataset based on two widely used text-based common knowledge QA datasets: TriviaQA\footnote{https://huggingface.co/datasets/mandarjoshi/trivia\_qa} and Natural Questions\footnote{https://huggingface.co/datasets/sentence-transformers/natural-questions}. 
We have verified that the training, validation, and test sets of TriviaQA have no overlap.
The data construction process involves the following three steps:
(1). Answer rewriting: The original answers in these datasets are typically short phrases. We use a language model via an online API to rewrite each answer into a complete sentence in a more conversational tone. This is done by prompting the model with both the original question and answer phrase, instructing it to produce a naturally phrased, full-sentence response.
(2). Speech synthesis: We apply zero-shot text-to-speech (TTS) to generate both the spoken questions and answers. To increase speaker diversity, we synthesize audio using thousands of different speaker prompts.
(3). ASR filtering: We run automatic speech recognition (ASR) on the generated speech and compute the word error rate (WER). QA pairs with an answer WER greater than 20\% are discarded to ensure transcription quality.

% \section{Experiments}

\section{Experimental setup}
\label{subsec:exp_setup}

\subsubsection{Training data}
\begin{table}[ht]
\centering
\caption{Statistics of training datasets. Each cell shows values for question/answer speech.}
\label{tab:qa_dataset_stats}
\begin{tabular}{lcc}
\toprule
\textbf{Training Dataset} & \textbf{Total Duration (hours)} & \textbf{Average Duration (s)} \\
\midrule
VoiceAssistant & 589.51 / 3422.36 & 4.53 / 26.39 \\
NaturalQuestions & 85.52 / 227.32 & 3.07 / 8.18 \\
TriviaQA & 207.40 / 179.58 & 5.40 / 4.82 \\
\bottomrule
\end{tabular}
\end{table}
Besides the two datasets curated in Section~\ref{subsec:data_curation}, we also include the VoiceAssistant\footnote{https://huggingface.co/datasets/gpt-omni/VoiceAssistant-400K} data for training.
We leverage the zero-shot TTS to generate the answer speech for the VoiceAssistant dataset.
The statistics of the three datasets are shown in Table~\ref{tab:qa_dataset_stats}.

\subsubsection{Model Configuration}
We use the open-source Phi4-MM\footnote{https://huggingface.co/microsoft/Phi-4-multimodal-instruct} speech encoder as our audio encoder. For the language model, For the language model, we adopt Phi4-MM 3.8B as the base model and further scale it up to 7B. 
The speech tokenizer\footnote{https://github.com/xingchensong/S3Tokenizer}, audio token vocoder (including both the flow matching model and neural vocoder) are directly adopted from CosyVoice 2~\cite{du2024cosyvoice}\footnote{https://github.com/FunAudioLLM/CosyVoice} without modification.

\subsubsection{Training details}
In line with the configuration described in the Phi4-MM paper~\cite{phi4mini}, we incorporate a Low-Rank Adaptation (LoRA)~\cite{hu2022lora} module with a rank of 320. Our experiments are conducted using two base language models, with 3.8 billion and 7 billion parameters, respectively. When equipped with LoRA, the total number of trainable parameters becomes 460 million and 707 million for the 3.8B and 7B models, respectively. Rather than updating the full model, we fine-tune only the LoRA layers.%, allowing the underlying language model to remain less affected.

\subsubsection{Evaluation details}
In the spoken question answering (SpokenQA) task, the model listens to a spoken question and generates a response that should contain the correct reference answer. Accuracy is measured by checking whether the reference answer appears in the model’s output.
We evaluate our system on three standard benchmarks: 
Llama Questions\footnote{https://huggingface.co/datasets/TwinkStart/llama-questions}~\cite{nachmani2024spoken}, 
Trivia-QA\footnote{https://huggingface.co/datasets/TwinkStart/speech-triavia-qa}~\cite{joshi2017triviaqa}, 
and Web Questions\footnote{https://huggingface.co/datasets/TwinkStart/speech-web-questions}~\cite{berant-etal-2013-semantic}.
It is worth noting that, there is no official test set for TriviaQA. Because the test samples may differ from those used in other papers, our results on TriviaQA are not directly comparable to open-source results and should be considered for reference only.

For evaluation, we report speech-to-text (S2T) accuracy, which measures whether the reference answer appears in the model’s text output.
We also evaluate speech-to-speech (S2S) accuracy by transcribing the model’s speech response using Whisper-large-v3~\cite{radford2023robust}\footnote{https://huggingface.co/openai/whisper-large-v3}, and checking whether the reference answer appears in the transcription.
To further assess alignment between the model’s speech and text outputs, we compute the S2S/S2T accuracy ratio, indicating how faithfully the speech output reflects the textual response.
In addition, we measure word error rate (WER) between the transcribed speech and the model-generated text, using Whisper-large-v3 for transcription. This provides an additional evaluation of consistency between the text and speech modalities.
The S2S/S2T accuracy ratio evaluates whether the model accurately verbalizes key answer-related content present in the text response, while the WER metric assesses how faithfully the generated speech reflects the entire textual output.
Our evaluation pipeline is based on MiniCPM-O\footnote{https://github.com/OpenBMB/UltraEval-Audio}.

\subsubsection{Baselines}
\begin{itemize}
    \item For the Thinker-Talker models—Minmo, and LLaMA-Omni 2.5~\cite{chen2025minmo, fang2025llama}—we report results from the Minmo paper, as the model weights are not publicly available. For LLaMA-Omni 2\footnote{https://github.com/ictnlp/LLaMA-Omni2}, we evaluate the models using their officially released checkpoints. We use the 3B, 7B, and 14B versions of LLaMA-Omni-2 as our main baselines due to their similar training setup—fine-tuned on only a few thousand speech QA data, unlike other models trained on much larger and more diverse datasets. Released in May 2025, LLaMA-Omni-2 also serves as a recent and relevant point of comparison.
    \item GLM-4-Voice~\citep{zeng2024glm} is among the first ones to adopt an interleaved decoding strategy for both pretraining and fine-tuning a Speech LM. Trained on millions of hours of audio, it supports real-time speech interaction by generating text and speech tokens in an alternating pattern with a fixed 13:26 ratio. The resulting speech tokens are passed through a flow matching model and subsequently a vocoder to synthesize the final waveform. We report results from MiniCPM-O\footnote{https://github.com/OpenBMB/MiniCPM-o}, as the numbers presented in the original GLM-4-Voice paper appear to underestimate the model’s actual performance. 
    \item Moshi~\cite{defossez2024moshi} is the first open-source speech LM that supports real-time speech-to-speech generation. It jointly models and decodes text and audio tokens with the parallel decoding pattern. We report results based on MiniCPM-O, as the original Moshi paper appears to underestimate the model’s speech output accuracy.
    \item GPT-4o~\cite{gpt4o} is a multimodal language model by OpenAI. We evaluate it by the Azure OpenAI platform\footnote{https://learn.microsoft.com/en-us/azure/ai-services/openai/}. 
    % Specifically, we select the ``gpt-4o-audio'' deployment with version 2025-01-01-preview.
\end{itemize}

\section{Experimental results}
\label{subsec:exp_result}
\begin{table*}[ht]
\centering
\caption{Evaluation results across LLaMA Questions, Trivia-QA, and Web Questions. ``Text'' and ``Speech'' denote accuracies for text and speech outputs respectively. ``Rel.'' is the ratio of ``Text out'' accuracy to ``Speech out'' accuracy.}
\vspace{-3pt}
\setlength{\tabcolsep}{4pt}
\renewcommand{\arraystretch}{1.2}
\begin{tabular}{l|ccc|ccc|ccc|c}
\toprule
\textbf{Model} & \multicolumn{3}{c|}{\textbf{LLaMA Questions}} & \multicolumn{3}{c|}{\textbf{Trivia-QA}} & \multicolumn{4}{c}{\textbf{Web Questions}} \\
 & \textbf{Text(\%) $\uparrow$} & \textbf{Speech(\%) $\uparrow$} & \textbf{Rel $\uparrow$} & \textbf{Text(\%) $\uparrow$} & \textbf{Speech(\%) $\uparrow$} & \textbf{Rel. $\uparrow$} & \textbf{Text(\%) $\uparrow$} & \textbf{Speech(\%) $\uparrow$} & \textbf{Rel. $\uparrow$} & \textbf{WER(\%) $\downarrow$} \\
\midrule
\rowcolor{modelblue}
(A1). GPT-4o & 84 & 74.67 & 0.89 & 73.73 & 65.92 & 0.89 & 51.97 & 48.23 & 0.93 & -- \\
\rowcolor{modelblue}
(A2). GLM-4-Voice (9B) & 64.7 & 50 & 0.77 & 39.1 & 36.4 & 0.93 & 32.2 & 32 & 0.99 & -- \\
\rowcolor{modelblue}
(A3). Moshi (7B) & 62.3 & 43.7 & 0.70 & 22.8 & 16.7 & 0.73 & 26.6 & 23.8 & 0.89 & -- \\
\rowcolor{modelblue}
(A4). Minmo (7B) & 78.9 & 64.1 & 0.81 & 48.3 & 37.5 & 0.78 & 55 & 39.9 & 0.73 & -- \\
\rowcolor{modelblue}
\midrule
(B1). LLaMA-Omni2-3B & 67 & 58.33 & 0.87 & 29.3 & 27.25 & 0.93 & 30.51 & 27.56 & 0.90 & 5.46 \\
\rowcolor{modelblue}
(B2). LLaMA-Omni2-7B & 72.33 & 64 & 0.88 & 36.33 & 32.71 & 0.90 & 33.17 & 30.07 & 0.91 & 5.69 \\
\rowcolor{modelblue}
(B3). LLaMA-Omni2-14B & 75 & 63.33 & 0.84 & 46.88 & 41.41 & 0.88 & 40.4 & 35.88 & 0.89 & 5.58 \\
\midrule
\rowcolor{modelorange}
(C0.1). Base model 3.8B & 71.67 & - & - & 34.57 & - & - & 36.22 & - & - & - \\
\rowcolor{modelorange}
(C0.2). Base model 7B & 77 & - & - & 43.75 & - & - & 41.38 & - & - & - \\
\midrule
\rowcolor{modelorange}
(C1). Ours 3.8B-ESI & 68.33 & 64.67 & 0.95 & 30.99 & 29.63 & 0.96 & 31.71 & 30.88 & 0.97 & 4.37 \\
\rowcolor{modelorange}
(C2). Ours 7B-ESI & 69.33 & 65 & 0.94 & 42.2 & 41.13 & 0.97 & 38.64 & 37.91 & 0.98 & 2.97 \\
% (C2). Phi-MM-ESI 7B & 70 & 65 & 0.93 & 45.81 & 43.76 & 0.96 & 40.51 & 39.18 & 0.97 & 3.25 \\

% Only for readout | Results for Enhance parallel
% \midrule
% \rowcolor{modelgreen}
% (D1). Phi4-MM-EP 3.8B & 69.67 & 65.33 & 0.94 & 36.06 & 34.41 & 0.95 & 32.55 & 31.27 & 0.96 & 5.22 \\
% \rowcolor{modelgreen}
% (D2). Phi4-MM-EP 7B & 70.33 & 64.67 & 0.92 & 45.32 & 42.79 & 0.94 & 40.41 & 38.25 & 0.95 & 5.13 \\
\bottomrule
\end{tabular}
\label{tab:speechqa_results}
\end{table*}

\begin{table*}[ht]
\centering
\caption{Comparison of decoding patterns using the same base model (Phi-MM 3.8B) and the same VoiceAssistant QA training data. ESI denotes our proposed early-stop interleaving strategy.}
\vspace{-3pt}
\setlength{\tabcolsep}{4pt}
\renewcommand{\arraystretch}{1.1}
\begin{tabular}{l|ccc|ccc|cccc}
\toprule
\textbf{Pattern} & \multicolumn{3}{c|}{\textbf{LLaMA Questions}} & \multicolumn{3}{c|}{\textbf{Trivia-QA}} & \multicolumn{4}{c}{\textbf{Web Questions}} \\
 & \textbf{Text(\%) $\uparrow$} & \textbf{Speech(\%) $\uparrow$} & \textbf{Rel. $\uparrow$} & \textbf{Text(\%) $\uparrow$} & \textbf{Speech(\%) $\uparrow$} & \textbf{Rel. $\uparrow$} & \textbf{Text(\%) $\uparrow$} & \textbf{Speech(\%) $\uparrow$} & \textbf{Rel. $\uparrow$} & \textbf{WER(\%) $\downarrow$} \\
\midrule
Interleave & \textbf{64.67} & \textbf{57.33} & 0.89 & 24.76 & 23.20 & 0.94 & 27.29 & 26.20 & 0.96 & 5.11 \\
Parallel   & 60.67  & 46.67  & 0.77  & 22.51  & 19.59  & 0.87  & 26.99  & 19.67  & 0.73  & 17.94  \\
ESI (Ours) & 63.67 & 55 & 0.86 & \textbf{27.29}  & \textbf{24.56}  & 0.90 & \textbf{28.96}  & \textbf{26.40}  & 0.91 & 5.72 \\
\bottomrule
\end{tabular}
\label{tab:decoding_pattern_results}
\end{table*}

\subsection{Main results}
We report main results as shown in Table~\ref{tab:speechqa_results}. (C0.1) and (C0.2) are the base language models, which serve as the initialization checkpoints for (C1) and (C2), and represent the top-lines for speech-in, text-out QA in their respective settings.
GPT-4o serves as a top-line reference with strong performance in speech-out and text-out accuracies.

Comparing our methods with models trained on millions of hours data:
(1). GLM-4-Voice is one of the earliest models to adopt the interleaved speech-text decoding strategy. Our proposed method, Phi4-MM with ESI, extends and improves this paradigm. On all benchmarks, Ours 7B-ESI consistently outperforms GLM-4-Voice, particularly in speech output accuracy and speech-to-text alignment (relative percentage).
Notably, Ours 3.8B-ESI surpasses GLM-4-Voice on the LLaMA Questions (and is on par on the Web Questions), despite using a significantly smaller base model (3.8B vs. 9B).
(2). Compared to Moshi, Ours 7B-ESI also achieves stronger and more stable performance across tasks. Even the smaller 3.8B variant (referred to as C1) outperforms Moshi in all metrics. 
(3). Our model C2 exceeds Minmo in both speech output accuracy and relative percentage, demonstrating more effective alignment between generated speech and text.
Note that our base models, C0.1 and C0.2, do not outperform Minmo on text-based QA. Minmo is with a better base model with outstanding intelligence than ours. We believe that replacing the base model could lead to significant performance improvements for our approach. % And we can pre-train our model with not only QA data to get better alignment.

Unlike models such as GLM-4-Voice, Moshi, and Minmo, which are trained on millions of diverse examples, LLaMA-Omni-2 and Phi4-MM with ESI are both fine-tuned using only a few thousand hours of QA samples. This makes them more comparable. 
When comparing Ours-3.8B ESI and 7B-ESI to the LLaMA-Omni-2 3B, 7B, and 14B counterparts:
(1). Phi4-MM with ESI shows clear advantages in speech-out accuracy, Rel \%, and WER. 
Take LLaMA Questions as an example, Ours 3.8B-ESI achieves 64.67\% speech-out accuracy, surpassing LLaMA-Omni2-3B (58.33\%) and even LLaMA-Omni2-7B (64.0\%). Its relative score of 0.95 further indicates a high degree of alignment between text and speech outputs. Moving to the 7B variant, Ours 7B-ESI pushes speech-out accuracy to 65\%, outperforming all LLaMA-Omni2 variants, including the much larger 14B model (63.33\%).
(2). When comparing models of similar size, Ours 3.8B-ESI outperforms LLaMA-Omni2-3B across all tasks. For example, On LLaMA-QA, it achieves 64.67\% speech-out accuracy versus 58.33\% for LLaMA-Omni2-3B, and shows stronger relative alignment (0.95 vs. 0.87). 
At the 7B scale, Ours 7B-ESI also consistently outperforms LLaMA-Omni2-7B for a large margin for all metrics on Web Questions and Trivia-QA.
on LLaMA questions, Ours 7B-ESI continues to outperform LLaMA-Omni2-7B, achieving 65\% vs. 64\% for the speech-out accuracy, and showing higher relative scores.

\subsection{Comparing different patterns}
As shown in Table~\ref{tab:decoding_pattern_results}, all three decoding strategies are evaluated in a controlled and fair comparison.

First, we observe that the interleaved decoding pattern consistently outperforms the parallel pattern across all datasets. For example, on LLaMA-QA, interleave achieves 57.33\% in speech-out accuracy versus 46.67\% for parallel; and in Web Questions, the relative alignment drops sharply from 0.96 (interleave) to 0.73 (parallel). In addition, the WER for the parallel pattern (17.94\%) is substantially higher than interleave (5.11\%), indicating degraded output speech-text alignment.

Our proposed ESI pattern achieves comparable or better results while reducing the total sequence length to approximately 75\% of that in standard interleaved decoding. Despite this, it matches or even surpasses interleaved one in performance: on Trivia-QA, ESI improves text-out from 24.76\% to 27.29\% and speech-out from 23.2\% to 24.56\%; on Web Questions, it also leads with 28.96\% text-out and 26.4\% speech-out.

\subsection{ESI pattern vs Interleaved pattern}

\begin{table}[ht]
\renewcommand{\arraystretch}{1.2}
\setlength{\tabcolsep}{2pt}
\centering
\caption{Performance comparison of decoding patterns (Interleave vs. Early-Stop Interleave).}
\begin{tabular}{l l | cc | cc | c}
\toprule
\textbf{Model} & \textbf{Dataset} & \multicolumn{2}{c|}{\textbf{LLaMA Questions}} & \multicolumn{2}{c|}{\textbf{Web Questions}} & \textbf{WER(\%)} \\
 &  & \textbf{Text(\%)} & \textbf{Speech(\%)} & \textbf{Text(\%)} & \textbf{Speech(\%)} & \\
\midrule
3.8B-I & V    & 64.67 & 57.33 & 27.29 & 26.20 & 5.11 \\
3.8B-ESI & V  & 63.67 & 55.00 & 28.96 & 26.40 & 5.72 \\
\midrule
3.8B-I & All  & 64.00 & 59.33 & 31.66 & 30.88 & 5.08 \\
3.8B-ESI & All & 68.33 & 64.67 & 31.71 & 30.88 & 4.37 \\
\midrule
7B-I & V     & 68.33 & 61.33 & 36.28 & 33.73 & 4.83 \\
7B-ESI & V   & 70.67 & 63.67 & 37.41 & 35.00 & 4.43 \\
\midrule
7B-I & All   & 71.67 & 65.67 & 36.68 & 35.10 & 2.92 \\
7B-ESI & All & 69.33 & 65.00 & 38.64 & 37.91 & 2.97 \\
\bottomrule
\end{tabular}
\label{tab:esi_results}
\end{table}

As shown in Table~\ref{tab:esi_results}, we compare interleaved decoding (I) and early-stop interleaved decoding (ESI) strategies using the same base models 3.8B and 7B on the LLaMA Questions and Web Questions benchmarks. In this table, ``V'' denotes training only on the VoiceAssistant QA dataset, while ``All'' includes additional data from Trivia-QA and NaturalQuestions. As results for Trivia-QA are consistent, we omit them here for limited spare space.

Across nearly all settings, ESI consistently improves performance or maintains comparable results while offering shorter token sequences during inference. For example:
On Web Questions, ESI improves speech-out accuracy from 26.2\% to 26.4\% (3.8B-V), and from 33.73\% to 35.00\% (7B-V).
On LLaMA Questions, switching from interleave to ESI improves text-out from 64.0\% to 68.33\% and speech-out from 59.33\% to 64.67\% for 3.8B-All.
For the 7B-V setting, ESI boosts text-out by 2.3\% and speech-out by 2.34\%.
These improvements highlight the effectiveness of the early-stop design in improving both accuracy while maintaining decoding.

\subsection{Dataset ablation study}
\begin{table}[ht]
\centering
\setlength{\tabcolsep}{3pt}
\caption{Effect of training data combinations on model performance (3.8B model with interleaving decoding). (a): VoiceAssistant, (b): Natural Questions, (c): Trivia-QA.}
\begin{tabular}{l|cc|cc|c}
\toprule
\textbf{Training Data} & \multicolumn{2}{c|}{\textbf{LLaMA Questions}} & \multicolumn{2}{c|}{\textbf{Web Questions}} & \textbf{WER(\%)} \\
 &  \textbf{Text(\%)} & \textbf{Speech(\%)} & \textbf{Text(\%)} & \textbf{Speech(\%)}  \\
\midrule
(a) & 64.67 & 57.33 & 27.29 & 26.20 & 5.11 \\
(a) + (b) & 61.00 & 56.00 & 30.24 & 29.30 & 4.20 \\
(a) + (c) & 62.33 & 56.67 & 28.52 & 28.32 & 6.18 \\
(a) + (b) + (c) & 64.00 & 59.33 & 31.66 & 30.88 & 5.08 \\
\bottomrule
\end{tabular}
\label{tab:data_ablation}
\end{table}

To investigate the effect of the generated datasets, (b) and (c), we compare the performance of models trained on different combinations of QA datasets as shown in Table~\ref{tab:data_ablation}. All experiments use the same 3.8B base model and interleaved decoding pattern. We omit the results for Trivia-QA (with the same trends as other datasets) due to space limitation.

Starting from (a), it uses only the VoiceAssistant dataset. 
When Natural Questions (b) or Trivia-QA is added in training, the model shows consistent gains on Web Questions—text-out and speech-out accuraties.
Combining all three datasets (a)+(b)+(c) leads to balanced and generally improved performance across all tasks. For example, LLaMA-QA speech-out increases to 59.33\%, and Web Questions achieves its best overall accuracy with 31.66\% text-out and 30.88\% speech-out.
These results show that the two generated datasets, (b) and (c), improve the overall performance.
\section{Conclusion}
\label{sec:conclusion}
In this work, we conduct a comprehensive and fair comparison on decoding strategies for joint speech-text generation within a unified language model. Our findings reveal that the interleaved pattern consistently outperforms its counterparts in both text and speech output quality.
To mitigate the inefficiency of interleaved decoding, we introduce a novel early-stop interleaved decoding pattern. This pattern allows the model to terminate text generation with a special token and proceed to generate only speech tokens, reducing sequence length by approximately 25\% on average. This not only improves inference speed but also preserves output quality and alignment, making it a practical solution for real-time conversations. Additionally, we demonstrate the importance of data quality in speech question answering. 
% By curating and augmenting QA datasets with diverse, well-structured examples, we significantly enhance the performance of speech-in, speech-out QA tasks across multiple benchmarks.
This paper provides practical insights for future speech LM development.

% \newpage
\bibliographystyle{IEEEtran}
\bibliography{IEEEabrv,refs}

\end{document}